\documentclass{iaus}
\usepackage{graphicx}
\usepackage{graphicx,bm,url}
\usepackage{natbib}

\newcommand{\EQ}{\begin{equation}}
\newcommand{\EN}{\end{equation}}
\newcommand{\EQA}{\begin{eqnarray}}
\newcommand{\ENA}{\end{eqnarray}}

{}
{}
{}

{}
{}
{}
{}
{}
{}
{}
{}
{}
{}
{}
{}
{}
{}
{}
{}
{}
{}
{}
{}

{}
{}
{}

{}

{}
{}

\newcommand{\yapj}[3]{ #1, {ApJ,} {#2}, #3}

\newcommand{\yapjl}[3]{ #1, {ApJ,} {#2}, #3}

\newcommand{\yan}[3]{ #1, {Astron.\ Nachr.,} {#2}, #3}

\newcommand{\yana}[3]{ #1, {A\&A,} {#2}, #3}

\newcommand{\yjetp}[3]{ #1, {Sov.\ Phys.\ JETP,} {#2}, #3}

\newcommand{\ymn}[3]{ #1, {MNRAS,} {#2}, #3}

\newcommand{\ypre}[3]{ #1, {Phys.\ Rev.\ E,} {#2}, #3}

\newcommand{\ybook}[3]{ #1, {#2} (#3)}

\newcommand{\pd}{\partial}
\title[Active longitudes and dynamos]
{Role of longitudinal activity complexes for solar and stellar dynamos}

\author[M.\ J.\ Mantere, P.\ J.\  K\"apyl\"a \& J.\ Pelt]
{Maarit J.\ Mantere$^{1}$, Petri J.\ K\"apyl\"a$^{1,2}$, Jaan Pelt$^{3}$}

\affiliation{
$^1$Department of Physics, PO Box 64, FI-00014 University of Helsinki, Finland\\
$^2$Nordita, KTH Royal Institute of Technology and Stockholm University, Roslagstullsbacken 23, SE-10691 Stockholm, Sweden\\
$^3$Tartu Observatory, T$\tilde{\rm o}$ravere, 61602, Estonia
}

\pubyear{2013}
\volume{294}
\pagerange{1--12}
\date{$ $Revision: 1.2 $ $}
\jname{Solar and Astrophysical Dynamos and Magnetic Activity}
\editors{A.G. Kosovichev, E.M. de Gouveia Dal Pino \& Y. Yan, eds.}
\begin{document}

\maketitle

\begin{abstract}
  In this paper we first discuss observational evidence of
  longitudinal concentrations of magnetic activity in the Sun and
  rapidly rotating late-type stars with outer convective
  envelopes. Scenarios arising from the idea of rotationally
  influenced anisotropic convective turbulence being the key physical
  process generating these structures are then presented and discussed
  - such effects include the turbulent dynamo mechanism, negative
  effective magnetic pressure instability (NEMPI) and hydrodynamical
  vortex instability. Finally, we discuss non-axisymmetric stellar
  mean-field dynamo models, the results obtained with them, and
  compare those with the observational information gathered up so
  far. We also present results from a pure $\alpha^2$ mean-field dynamo
  model, which show that time-dependent behavior of the dynamo
  solutions can occur both in the form of an azimuthal dynamo wave
  and/or oscillatory behavior related to the alternating energy levels
  of the active longitudes. 
\end{abstract}

\firstsection

\section{Introduction}

The spectroscopic time series obtained for rapidly rotating late-type
stars, analyzed with Doppler imaging methods, have revealed huge high-
latitude temperature anomalies unevenly distributed over the stellar
longitudes. The analysis of photometric time series have in many cases
confirmed the spectroscopic results, and allowed the follow-up of the
evolution of the light curve minima for longer time intervals and with
better time sampling. From these investigations it seems clear that in
many stars the magnetic activity manifests itself as active
longitude(s) that undergo apparently quite irregular phase jumps,
sometimes called flip-flops, and drifts in the rotational frame of
reference. In terms of dynamo theory, a change from the solar-type
mostly axisymmetric and quite regular oscillatory solution to a
non-axisymmetric non-stationary mode is expected to occur as the
rotation rate is increasing.

Some of the phenomena related to these active longitudes can rather
straightforwardly be understood using mean-field dynamo theory (see
e.g. Krause \& R\"adler, 1980). The change from axi- to
non-axisymmetric solutions with increasing rotation rate, for
instance, is a prediction dating back to kinematic linear dynamo
solutions, that has been later confirmed by more complex non-linear
modeling. Time-dependent solutions, with the angular frequency of the
non-axisymmetric modes being different from the axisymmetric mode and
stellar rotation constituting an azimuthal dynamo wave, have also been
recognized as common linearly preferred and non-linearly stable
solutions to the dynamo equations. Understanding and explaining the
flip-flops has remained a prominent challenge both observationally and
theoretically. Photometry and Doppler imaging naturally do not give
the information of the polarity of the spots on the active
longitudes. Therefore it has been impossible to differentiate whether
the phase jumps are related only to the alternation of the magnetic
energy levels of the spots, or are polarity changes also related to
the phenomenon. Only very recently such observational data has become
available; as will be discussed in Sect.~\ref{sect:obs}, the analysis
and interpretation of the data is far from straightforward, and clear
conclusions are still missing. Theoretically, oscillatory solutions
are more generally associated with $\alpha \Omega$ dynamo solutions,
where differential rotation is dominating the field generation over
the $\alpha$ effect. In the rapid rotation regime, however,
differential rotation is expected to be suppressed. It has remained
challenging to find oscillatory solutions of pure $\alpha^2$ type,
although it has been known since the kinematic calculations that such
solutions are possible. Recently, more results of oscillatory
$\alpha^2$ dynamo solutions have emerged, and in this paper we also
demonstrate one such model in Sect.~\ref{subsect:pc_dynamo}.

One of the most puzzling results related to the spectropolarimetric
observations, yielding the surface magnetic field strength and
orientation with the surface temperature distribution, is that very
seldom clear anticorrelation between the magnetic field strength and
temperature (see e.g. Kochukhov et al.\ 2012 and references therein),
expected from the sunspot analogy, is found. From theoretical point of
view, it is, indeed, quite poorly understood how the dynamo-generated
sub-surface fields transform themselves into sun- or starspots, and
whether the same mechanisms are actually at play in all the objects
under study. The sunspot sizes, for instance, are on average much
smaller than the scale of the toroidal magnetic field belt that they
emerge according to dynamo theory - this is to be contrasted to the
starspots that have surfaces areas large enough to be regarded as
global large-scale phenomena as the dynamo-generated field itself.

The rather widely accepted paradigm of sun- and starspot formation is
the so called rising flux tube model (see e.g. Choudhuri \& Gilman,
1987), according to which strong magnetic flux is generated and stored
in the tachoclinic shear layer just beneath the convectively unstable
region. The magnetic field becomes unstable and rises to the surface,
in the form of thin flux tubes, due to buoyancy force, during which
process it becomes only weakly affected by the turbulent motions in
the convection zone. Closely related to this paradigm is the flux
transport dynamo scenario (see e.g. Dikpati \& Charbonneau, 1999),
furthermore relying on the inductive effect due to sunspot decay near
the solar surface (Babcock-Leighton effect) accompanied with a
conveyor belt provided by meridional circulation, reviewed by
Choudhuri (2013) in these proceedings. Again, convective turbulence
plays a negligible role in the flux transport model, even the
turbulent diffusion being reduced significantly from the simple
estimates e.g. from mixing length theory. In this paper, we
concentrate more on the turbulent dynamo picture (see e.g. Brandenburg
\& Subramanian, 2005; Pipin, 2013 in these proceedings), which relies
on the idea of magnetic field generation both by rotationally affected
anisotropic convective turbulence and large-scale flows and
non-uniformities in the stellar rotation profile. Sun- and starspot
formation mechanisms directly related to turbulence effects include
the negative effective magnetic pressure instability (NEMPI) and
hydrodynamic vortex instability, both of which are also reviewed in
this manuscript. In the context of NEMPI from turbulent convection, we
also refer to the paper by K\"apyl\"a et al. (2013) in these
proceedings.

Next, we give a short review of stellar dynamo models capable of
solving for non-axisymmetric modes, and the results obtained with them
in Sect.~\ref{sect:dynamos}. We note that results in this field are
much less abundant than in the solar case, where axisymmetric (in two
dimensions) modeling is adequate. Finally, a summary and main
conclusions are presented in Sect.~\ref{sect:discussion}.

\section{Observational studies of active longitudes}

The study of longitudinal concentrations of magnetic activity started
with the accumulation of data of the distribution of solar activity
tracers on its surface (systematic sunspot observations, solar flares,
and so on). It was realized that they do not occur completely randomly
over the longitudes, but `hot spots' or `activity nests' surviving
over several years, maximally over the whole solar cycle, in which the
major sunspots and flares were accumulating over time, frequently
occur. The first tool was to create family trees (time-longitude
diagrams) of the activity complexes, aiming at following the pattern
their formed over time on the solar surface - using this technique it
soon became evident that the hot spots were not always moving with the
same rotation rate as the solar surface, nor the Northern vs Southern
activity complexes as a synchronized structure (see e.g. Bai, 1987 and
references therein). Evidently, something interesting was going on,
which ignited an intensive study of the phenomenon with sophisticated
time series analysis tools.

\subsection{Is the solar magnetic field axisymmetric?}

Several parametric and non-parametric time series analysis methods
have been used to analyze the distribution of sunspots and solar
flares over the past few decades. As the family tree analysis had
already revealed that the rotation period of the hot spots was
unlikely to match with the surface rotation rate, the first hypothesis
was to assume that it does not, and try to find the most suitable
rotation period that produces the most uneven distribution (i.e. the
strongest clustering) of the activity traces over the solar
longitudes. Again, South and North seemed to be decoupled, the other
natural hypothesis being to keep them separate in the analysis. Very
often the term 'active longitude' is used, in a strict sense meaning,
that at any latitude, being it Northern or Southern, the activity
should manifest itself at the same longitude. This definition does not
make much sense in any observational case - the solar hemispheres
behave in a decoupled manner, and usually only a fraction of the whole
latitudinal extent is seen of other stars. In the dynamo models,
however, the solutions indeed form a coherent longitudinal structure
over the whole latitudinal range, fulfilling the definition of an
active longitude. Nevertheless, the paradigm of active longitudes has
strongly influenced the discussion and analysis methods of solar
activity tracers; very often bimodal distributions are the ones
searched for, i.e. the distributions producing the strongest
two-peaked distributions are normally ranked to be the most
desirable. Even if not, the statistical significance of longitudinal
clustering of {\it any} number of peaks in the histograms has been
found to be very low. Furthermore, the best period calculated for the
whole data set might not represent well both hemispheres, nor be the
best if the data was divided into shorter intervals. This has led to
the overall conclusion that the non-axisymmetric component, at least
the one that would be rigidly rotating, of the solar dynamo must be
very weak and apparently dynamically changing over time.

It is well known that the solar convection zone is differentially
rotating so that at the surface, the pole rotates slower than the
equator. The next hypothesis adopted was to postulate, that the
surface distribution of activity traces becomes affected by this
rotation pattern. This essentially means that time series analysis
methods allowing for one free parameter more to be fitted, namely the
amount and direction of the surface differential rotation pattern,
were developed. Again, using the maximal clustering as the measure of
the statistical significance of the solution, it was claimed
(Berdyugina \& Usoskin, 2003; Usoskin et al., 2005; Berdyugina et al.,
2006) that a persistent system of active longitudes, comprising
roughly 10\% of the total number of sunspots, was present in the Sun,
made visible by the `passing' sunspot distribution on the surface
experiencing differential rotation. The existence of this stroboscopic
effect was put under doubt by the following investigations (Pelt et
al.\ 2005; 2006), which showed that even the differentially rotating
constructions produced longitudinal distributions that had a
significance level comparable to a statistical fluctuation. On the
other hand, it was shown (Pelt et al.\ 2010) that the sunspot
distribution, and therefore the magnetic field structure from which it
originates, is indeed affected by differential rotation. Moreover, it
was shown that nonparametric methods can be used to determine the
coherence time of the non-axisymmetric structures; the average
coherence time was estimated to be roughly 10-15 Carrington rotations.

As a conclusion, one might say that a weak non-axisymmetric component
of the solar magnetic field exists, and is affected by the
differential rotation of the solar convection zone. Due to its
weakness and dynamical nature, it is very hard to quantify it even
with the best statistical tools and despite of the long time span of
the sunspot data. Most of the solar dynamo models work under the
assumption of axisymmetry, which seems to be safe, taken what is said
above. Here we note, however, that also non-axisymmetric solar dynamo
models have been developed (e.g. Moss, 1999), where the excitation of
a sub-dominant non-axisymmetric magnetic field modes has been seen. These
models are kinematic, i.e. no back reaction of the magnetic field on
the velocity field is taken into account. Therefore, these models do
not properly address the dynamical significance of these modes.

\begin{figure}[t!]\begin{center}
\includegraphics[width=0.75\textwidth]{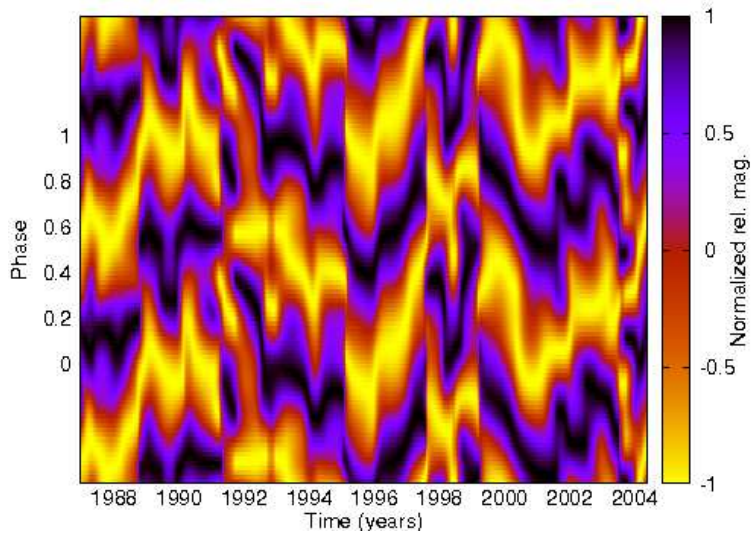}\\
\end{center}\caption[]{Photometric lightcurve of FK~Comae Berenices
  for 1987-2004 analyzed with the Carrier Fit method (Pelt et al.,
  2011; see also Hackman et al., 2012b). Number of model harmonics used
  is $K = 3$, number of modulation harmonics $L=20$, modulation period
  of 9000 days, and carrier period
  $P_0$=2.4002466d.}\label{fig:fkcom}\end{figure}

\subsection{Rapidly rotating late-type stars}\label{sect:obs}

Direct observational ways to detect and characterize spots and their
longitudinal distributions on stellar surfaces include photometric
lightcurve analysis, spectroscopic Doppler imaging, and
spectropolarimetric Zeeman-Doppler imaging. Observational result that
has had the strongest impact on the discussion of active longitudes
came from the analysis of the photometry of FK Comae Berenices by
Jetsu et al.\ (1993). By analyzing 25 years of photometry, two active
longitudes, roughly 180 degrees apart were found. The activity was
seen to abruptly switch from one longitude to the other, the jump
detected three times in the data used. This phenomenon was given the
name `flip-flop'; analysis of lightcurves of other rapidly rotating
late-type stars (e.g. Berdyugina \& Tuominen, 1998) soon established
the fact that flip-flops are a common phenomenon related to the
stellar cycle. Doppler imaging results, yielding the surface
temperature distribution, confirmed the photometric results
(e.g. Korhonen et al., 2004 and references therein), although with
poorer time sampling.

When these stars were followed up for a longer time span, it first
became evident that the active longitude system is not stable in the
rotational frame of reference of the stellar surface, i.e. the
non-axisymmetric system either formed straight, inclined, railroads
either upwards and downwards on the phase-time plots (e.g. Berdyugina
\& Tuominen, 1998), or even more complicated patterns of railroads
with changing inclination (Jetsu et al., 1999 for V~1794~Cyg),
sinusoidal paths (Berdyugina et al.\ 2002 for LQ~Hya; Berdyugina \&
J\"arvinen, 2005 for AB~Doradus) and so on. This revealed the other type
of time-dependent behavior of the active longitudes - the
non-axisymmetric system rotates with a different speed than the star
itself, and this rotation period actually changes over time. This does
not sound completely unfamiliar, as the solar hot spots are known to
exhibit the same behavior. Secondly, the longer the stars were
followed, the less periodic and regular the flip-flop events
appeared. To illustrate the situation for FK Comae Berenices, we have
re-analysed its photometry for the years 1987-2004 (Ol\'ah et
al., 2006; Korhonen et al., 2007) with the Carrier Fit (CF) method (Pelt et
al., 2011), see Fig.~\ref{fig:fkcom}. Here, we have used the
well-known photometric rotation period of $P_0\approx$ 2.4d as the
carrier in our analysis, and computed the continuous longitudinal
distribution of spots, appearing as dark regions in the plot, on the
star. One of the 'well-behaved' flip-flop events reported by Jetsu et
al.\ (1993) is visible during the first 3 years of the data. Since
then, at least four other flip-flop type events have occurred, but the
time intervals between them is clearly not regular. Moreover, railroad
patterns (tracks tilted up- and downwards) appear in the data since
1995 indicating that the non-axisymmetric structure does no longer
rotate with the photometric rotation period, but can lag behind or
speed up with respect to it. This photometric analysis is in rough
agreement with Doppler imaging results, see Korhonen et al.\
(2004). For a more detailed analysis of the phase jumps we refer to
Ol\'ah et al.\ (2006) and Hackman et al.\ (2012b).

\begin{figure}[t!]\begin{center}
\includegraphics[width=0.75\textwidth]{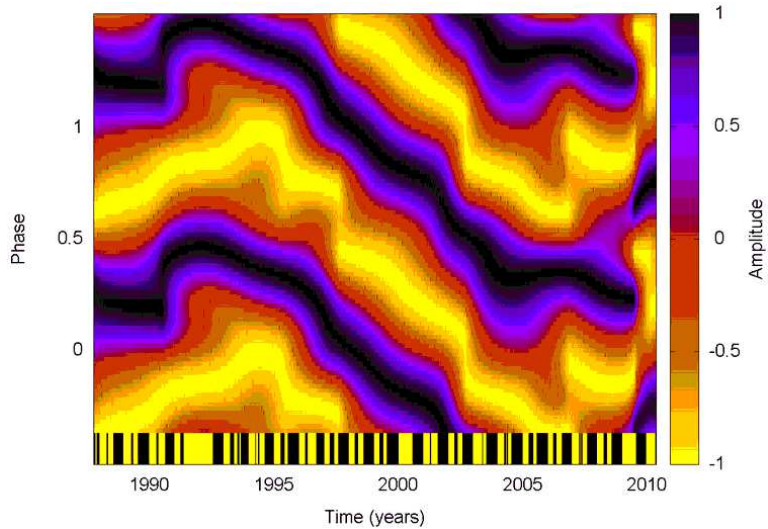}
\end{center}\caption[]{Photometric lightcurve of II~Peg for 1988-2010
  analyzed with the Carrier Fit method (Pelt et al., 2011). Number of
  model harmonics used is $K = 3$, number of modulation harmonics
  $L=5$, modulation period of 10000 days, and Carrier period
  $P_0$=6.724333d.}\label{fig:iipeg}\end{figure}

The railroad pattern is especially clear and persistent in the case of
the RS CVn binary system's primary giant component II~Peg (see
Fig.~\ref{fig:iipeg}; Lindborg et al., 2013). Instead of two active
longitudes, in this star we see the dominance of one dark region, with
sporadic appearance of a secondary minimum roughly 180 degrees apart
from the primary one (see Berdyugina et al., 1998, 1999; Lindborg et
al., 2011; Hackman et al., 2012a for surface images). The dominance of
the other active longitude could be thought to arise from the
binarity, but the situation is far from being that simple. Namely, the
CF analysis reveals a persistent downward trend in the phase-time
diagram over most of the timespan, indicating that the spotted
structure on a single active longitude is rotating faster than the
star itself. This is a very peculiar finding, especially reflected
upon the fact that in the system of close binary stars, the orbital
period becomes synchronized. It is possible that the synchronization
of the orbit has not yet fully occurred, but then the dominance of one
active longitude due to binarity and synchronism appears
invalid. Moreover, the systematic trend seems to break down towards the
end of the data set. During this epoch, Doppler imaging (Hackman et
al., 2012a) and Zeeman Doppler imaging (Kochukhov et al., 2012) suggest
that the magnetic activity level of the star is attaining a
minimum. Generally, the analysis of photometry is in good agreement
with the Doppler imaging results (see e.g. Hackman et al., 2011,
Lindborg et al., 2013), although the decreasing trend in the spot
intensity is not visible in Fig.~\ref{fig:iipeg} due to the adopted
normalization in our plot (each stripe is normalized with its extrema
to enhance the drift pattern during the epoch of weaker spots).

Comparing photometry and Doppler images to magnetic maps (Kochukhov et
al., 2012) brings further complications into the interpretation,
although it has to be kept in mind that magnetic maps exist only for
the lower state activity epoch - magnetic maps clearly show two active
longitudes of different polarities virtually at all times the star has
been imaged, which undergo polarity changes without any signs of
flip-flop occurring with the other methods. Only exception to this are the
most recent observing seasons (2008-2010), during which a polarity
change might be associated with the abrupt-looking phase behavior seen
also in Fig.~\ref{fig:iipeg}.

\section{Explanations in the framework of turbulent dynamo theory}

To understand the gradually building up observational picture, theory
should address at least the following two fundamental issues: Firstly,
the models should be capable of reproducing the transition from
oscillatory, mostly axisymmetric dynamo in the $\alpha \Omega$ regime
(differential rotation dominating over the effect of convective
turbulence) into the dominantly non-axisymmetric dynamo in the pure
$\alpha^2$ regime (field generation presumably taking place with
negligible differential rotation) when the rotation rate of the star
is increased. The usage of 'presumably' here refers to the fact that
we do not actually yet observationally know the internal rotation
profiles for other stars than the Sun; the only information comes from
modeling (see e.g. Kitchatinov \& R\"udiger, 1999). Secondly, what is
the relation of the large-scale dynamo-generated magnetic field and
the sun- and starspots themselves, and how does one form spots from
the underlying dynamo field?

Both of these issues have been addressed from the point of view of a
turbulent dynamo, which viewpoint is adopted in this paper; the
competing theory of the transport of rising flux tubes is discussed in
length by Choudhuri (2013) in these proceedings. According to the
theory of a turbulent dynamo (for an extensive review, see Brandenburg
\& Subramanian, 2005), the dynamo instability itself is affected by
the presence of turbulence so that a collective inductive action plus
an enhanced diffusion is caused by turbulence effects; as the
convection zone is turbulent throughout, these effects are distributed
over the total extent of it - sometimes the term distributed dynamo is
used to denote the turbulent dynamo process. Turbulence effects can
also lead to certain types of instabilities, that have been shown to
result in the formation of magnetic field concentrations or
global-scale temperature anomalies. On the other hand, structure
formation in convectively turbulent flow can occur via other, less
specified, routes, which provides us a starting point for the
theoretical discussion.

\begin{figure}[t!]\begin{center}
\includegraphics[width=0.32\textwidth]{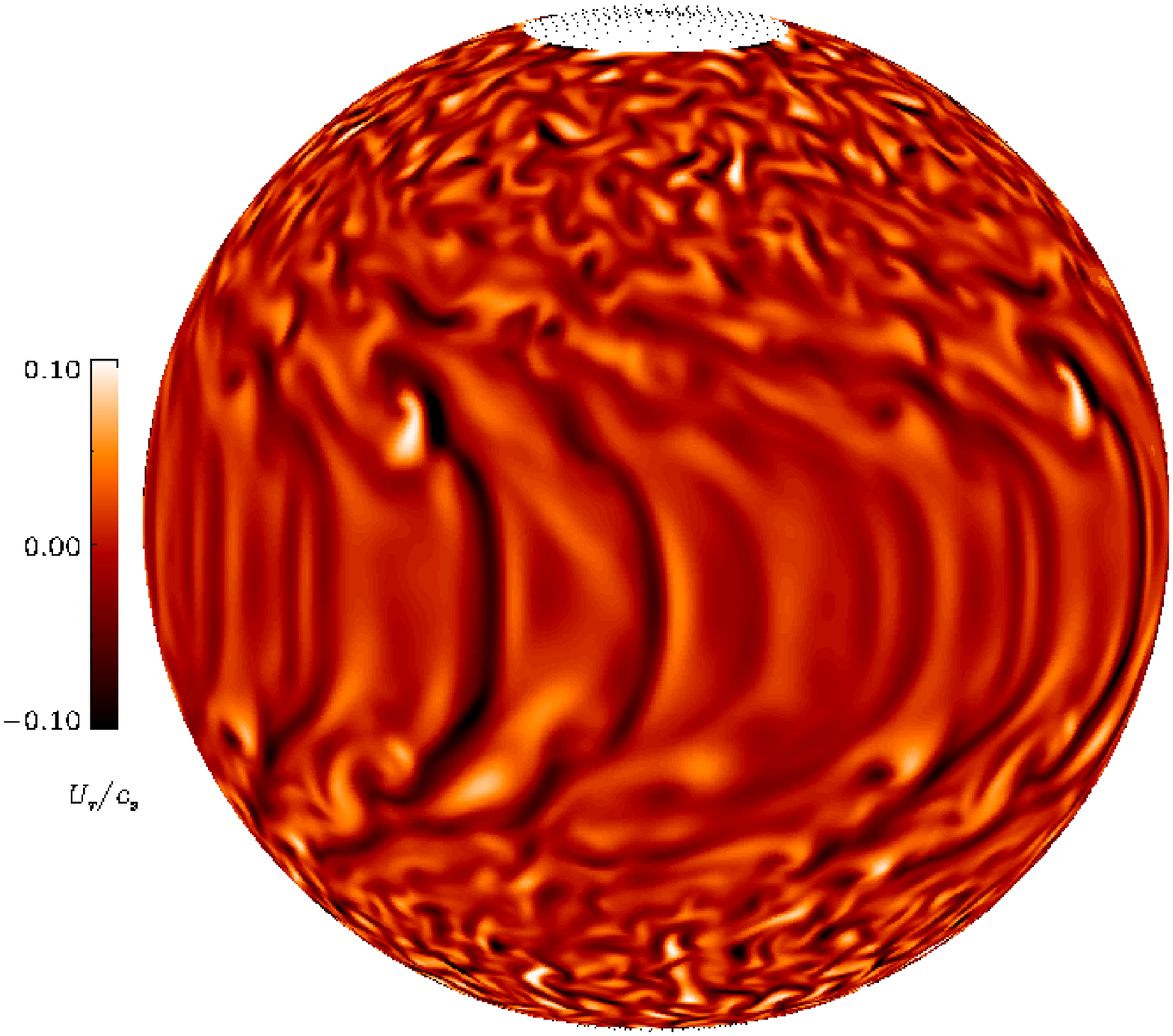}
\includegraphics[width=0.32\textwidth]{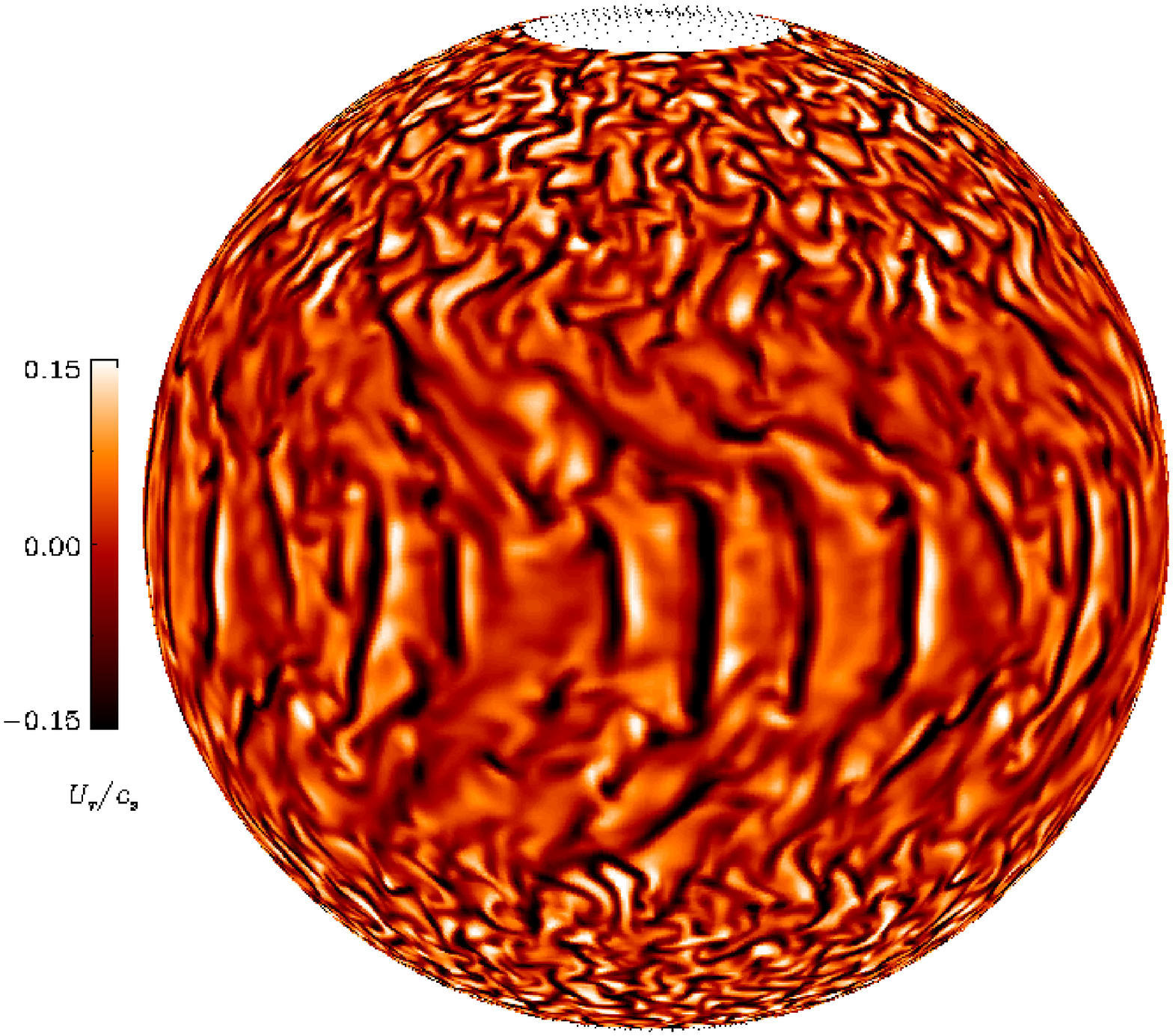}
\includegraphics[width=0.32\textwidth]{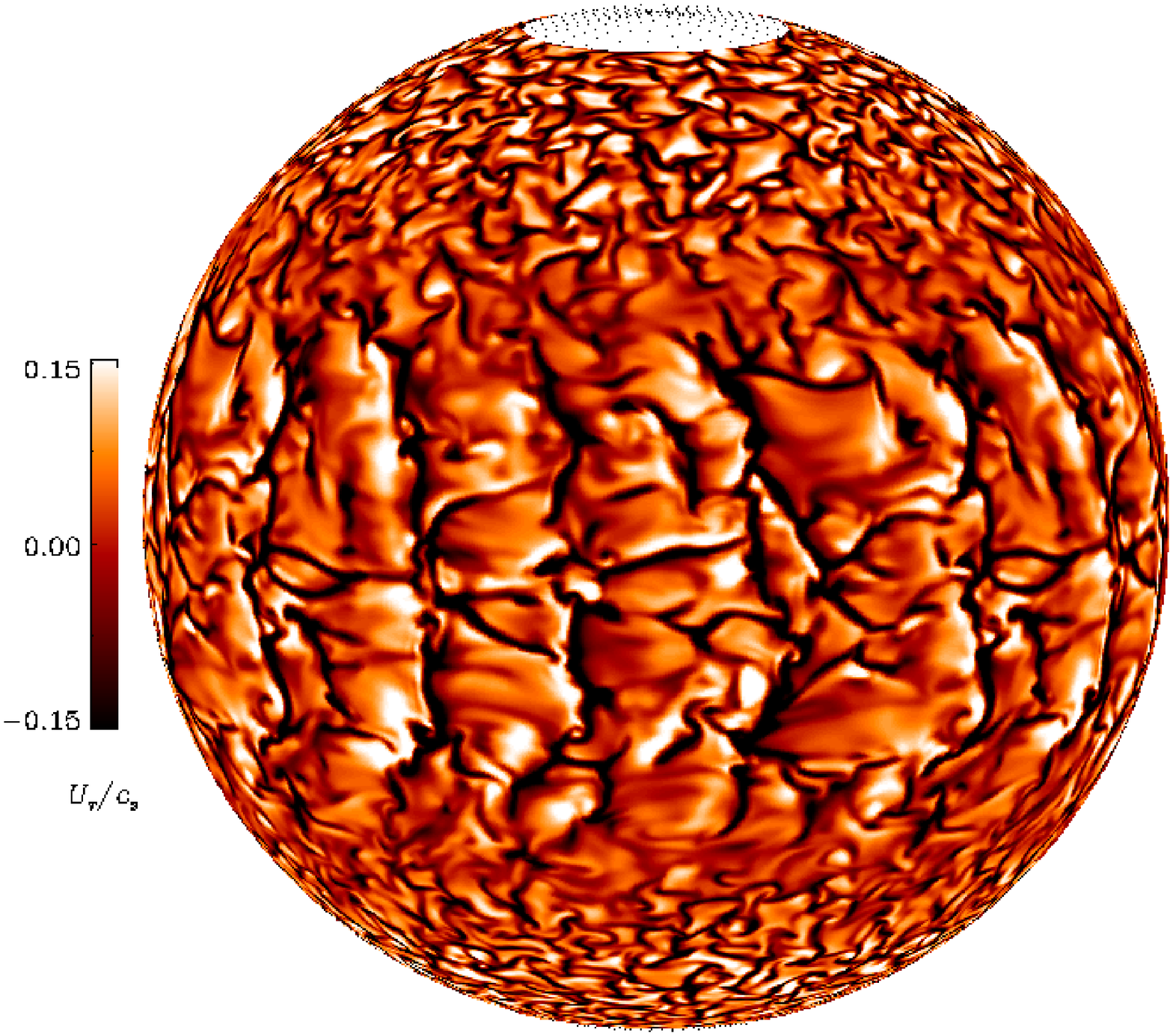}
\end{center}\caption[]{Radial velocity $U_r$ from three simulations of
  rapidly rotating turbulent convection with density contrasts of 5
  (left), 30 (middle), and 100 (right) in spherical wedge
  geometry. Adapted from \cite{KMB11}.}\label{fig:balls}\end{figure}

\subsection{Spontaneous formation of magnetic field concentrations in turbulent convection simulations}

Direct numerical solutions of magnetized convection have reached a
level of sophistication where the formation of flux concentrations can
be self-consistently modeled. Global and semi-global simulations of
rapidly rotating convection exhibit non-axisymmetric convection at low
latitudes when the magnetic field is weak or absent \cite[e.g.][see
the leftmost panel of
Fig.~\ref{fig:balls}]{BBBMT08,KMB11,ABBMT12}. This behavior is most
prominent near the onset of convection. Furthermore, increasing
density stratification also seems to suppress the non-axisymmetric
modes \citep[][compare the middle and rightmost panel of
Fig.~\ref{fig:balls} to the leftmost one]{KMB11}. However, as the
magnetic fields due to dynamo action become dynamically important, the
non-axisymmetric convection pattern disappears and the large-scale
magnetic fields are also mostly axisymmetric \citep[e.g.][]{BMBBT11};
see, however, Miesch et al. (2013) in these proceedings.

Another line of study concerns local simulations where some sort of
magnetic field is imposed or advected into the system. In these cases
turbulent convection can rearrange the magnetic field
self-consistently into concentrations in contrast to the sunspot
simulations of Rempel et al.\ (2009) where the spots are the result of
an imposed magnetic field structure. The outcomes of these models are
either large-scale magnetic structures (e.g. Tao et al., 1998), or can
resemble pores (Kitiashvili et al., 2010) or even bipolar regions in
the Sun (Stein \& Nordlund, 2012). The exact mechanism responsible for
the flux concentrations in these cases is not yet certain.

\subsection{NEMPI}

A possible mechanism for generating flux concentrations that
ultimately lead to the formation of sunspot and starspots is due to a
negative contribution of turbulence to the effective magnetic pressure
\citep[e.g.][]{KRR90,RK07}, a.k.a.\ the negative effective magnetic
pressure instability (NEMPI). This effect can form flux concentrations
even from uniform, sub-equipartition, magnetic fields.

Mean-field models and direct numerical simulations of forced
turbulence have established that the negative effect exists
\citep[e.g.][]{BKR10,BKKR12}, and the instability itself has been
detected in direct numerical simulations (DNS) of forced turbulence
\citep{BKKMR11}. A negative effect on magnetic pressure has also been
found for convection \citep{KBKMR12} but no NEMPI has been detected so
far (see, however, K\"apyl\"a et al.\ (2013) in these proceedings).

\subsection{Large-scale hydrodynamic vortex instability}

\begin{figure}[t!]\begin{center}
\includegraphics[width=5cm]{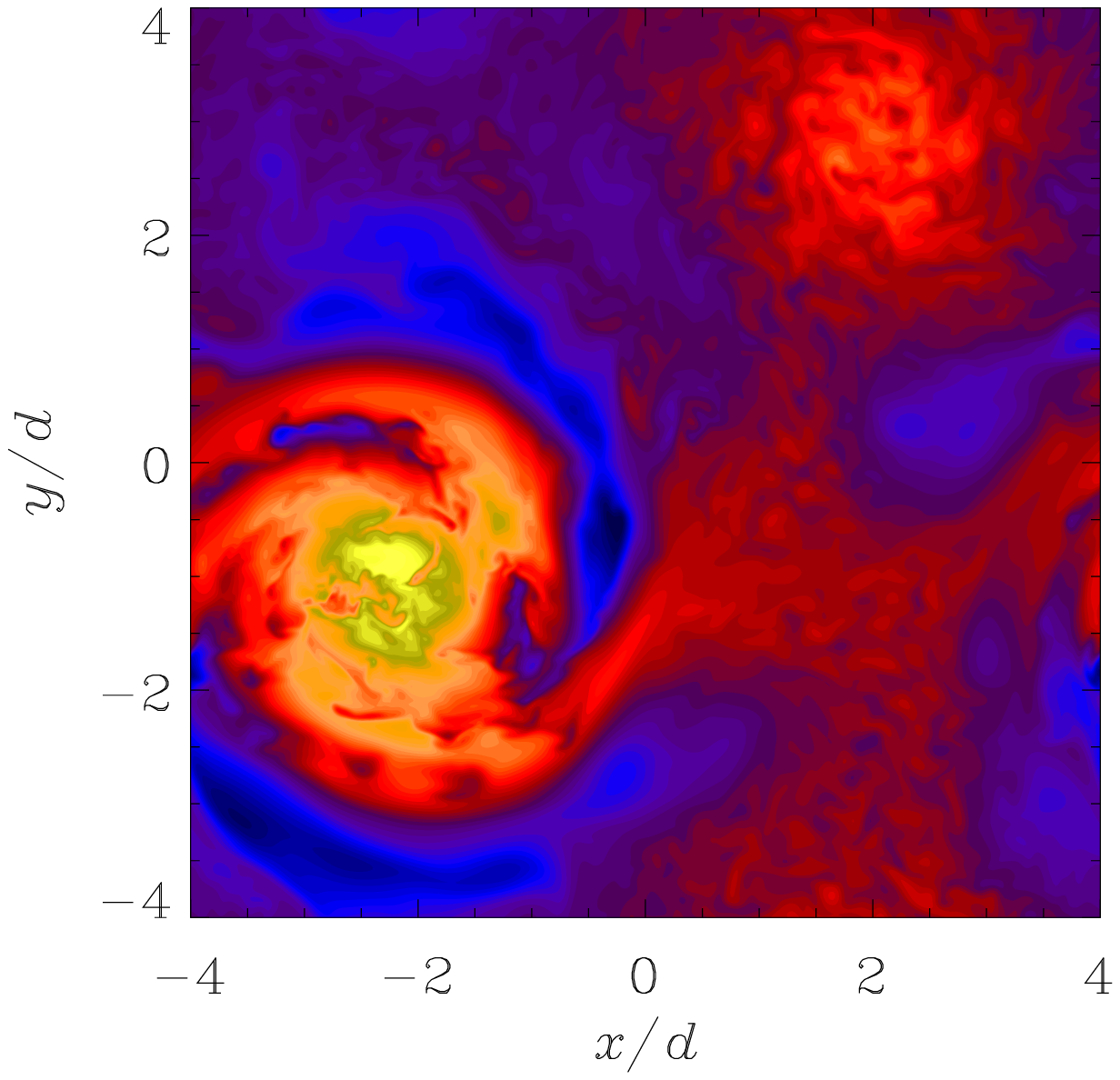}\hspace*{-1.cm}
\includegraphics[width=5cm]{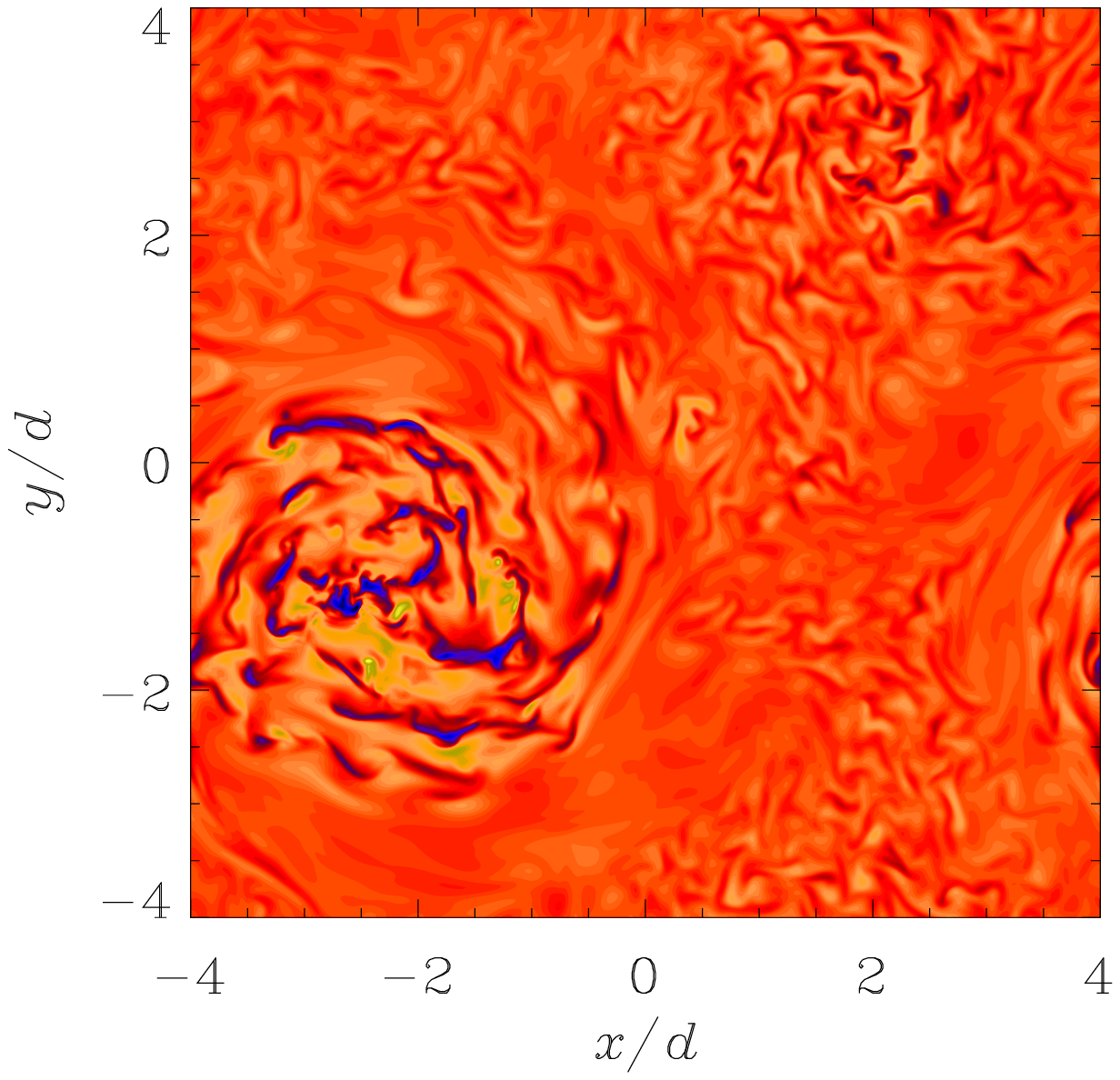}\hspace*{-1.cm}
\includegraphics[width=5cm]{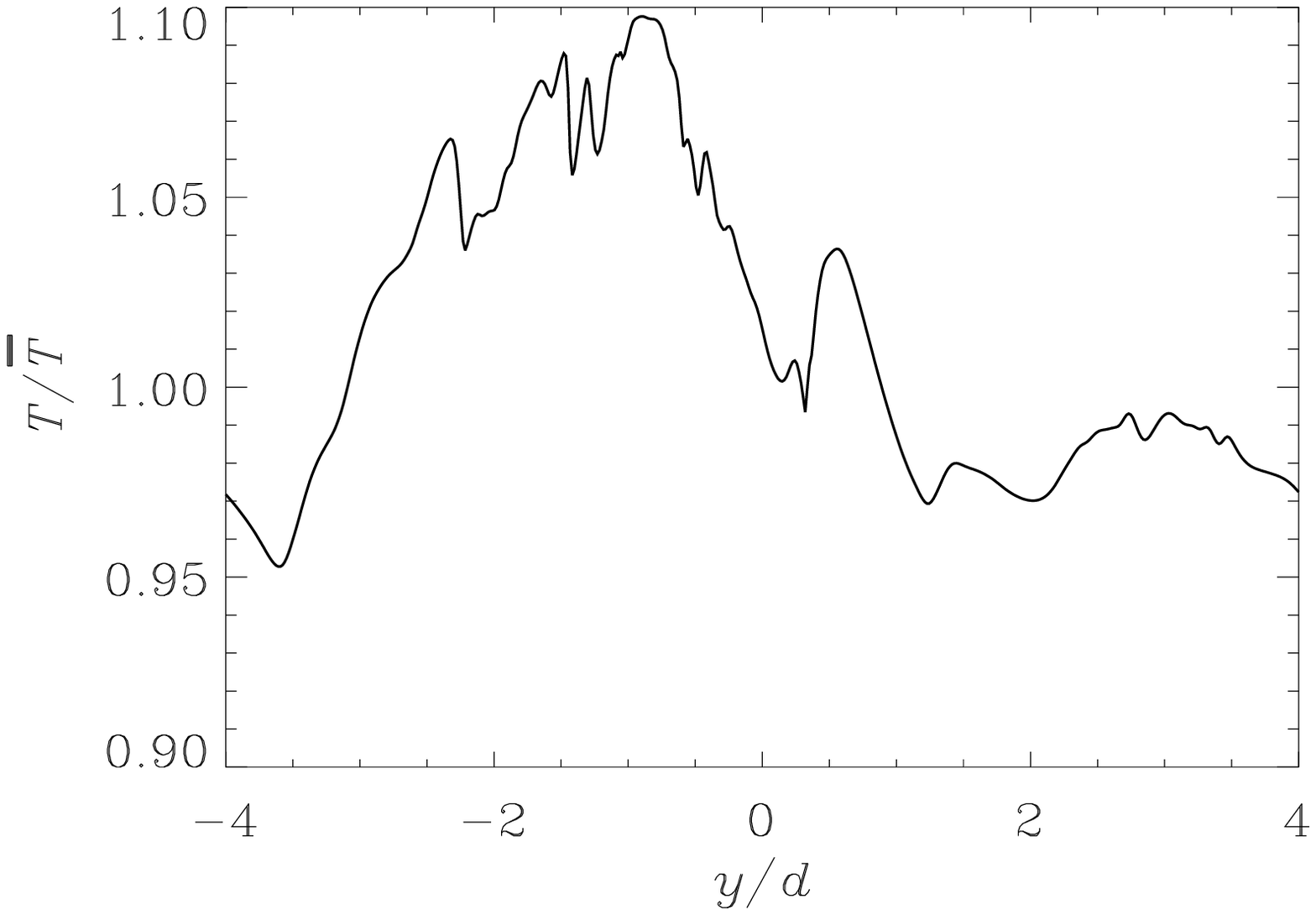}
\end{center}\caption[]{Large-scale vortices in rapidly rotating
  turbulent convection in Cartesian geometry. Left and middle panels
  show the temperature and vertical velocity, respectively, near the
  surface of the convectively unstable region. The right panel shows a
  cut of the temperature through the warm spot in the lower left
  corner of the left panel. Adapted from
  \cite{MKH11}.}\label{fig:vortex}\end{figure}

The apparent uncorrelation of magnetic fields and temperature
anomalies in Zeeman--Doppler maps of stars is a puzzling feature,
which may be attributable to the limited spatial resolution of the
current ZDI maps (see discussion in Kochukhov et al., 2012), or
explained with a new mechanism by which starspots can be generated
without a correlation of magnetic fields and temperature. The latter
possibility has been recently considered on the basis of results from
local \emph{hydrodynamic} simulations of rapidly rotating turbulent
convection \citep{KMH11,MKH11}. It turns out that when certain
threshold values of the the Reynolds and Coriolis numbers, describing
the effects of molecular viscosity and rotation on the flow, are
exceeded, large-scale vortices appear in the system. When rotation is
gradually increased, cool anti-clockwise vortices appear first. For
more rapid rotation also warm clockwise cyclones appear (see
Fig.~\ref{fig:vortex}). The temperature anomaly in the cyclones is at
least ten per cent of the surface temperature which similar to the
spots seen in Doppler images of rapidly rotating stars. Similar
vortices have not yet been found in global or semi-global
simulations. A probable reason is too low resolution achievable at the
moment. Furthermore, the vortices also seem to disappear in
simulations where a magnetic dynamo is present
\citep{KMB12a}. Currently it is unclear whether the cyclones can
coexist with strong magnetic fields.

\subsection{Non-axisymmetric mean-field dynamos}\label{sect:dynamos}

Modeling of the solar dynamo is at the verge of becoming accessible by
DNS (Ghizaru et al.\ 2010), the first solar-type solutions
having been obtained from such models (K\"apyl\"a et al., 2012c). Such
computations, however, are computationally expensive, and mean-field
modeling still has clear advantages in this repspect. The solar
mean-field models usually work under the assumption of axisymmetry,
reducing the problem into two spatial dimensions, whereas it is
obvious from observations that this assumption is not valid for the
more rapidly rotating stars with dominantly non-axisymmetric surface
temperature and magnetic field configurations. Non-axisymmetric models
necessarily solve for the azimuthal dependence of the physical
quantities, either on a finite-difference or -volume grid, or using
spherical harmonic decomposition and Fourier transform.

The non-axisymmetric solutions to the kinematic $\alpha^2$ equations
in simple setups were computed already decades ago (see e.g. R\"adler,
1975; Krause \& R\"adler, 1980), providing more or less all the
essential properties of the solutions: when rotation rate is increased
high enough, the preferentially excited modes are the non-axisymmetric
ones, typically representing quadrupolar symmetry (S1). The solutions
are usually non-oscillatory, but still time-dependent, as the modes
migrate either east- or westward. The nonlinear, sometimes even
dynamical, models developed since the 1990's (see e.g. Moss et
al. 1995; K\"uker \& R\"udiger 1999) have shown that also
antisymmetric dipolar solutions (A1) can appear as the nonlinearly
stable solutions (Tuominen et al.\ 2002). In this type of a solution,
two high-latitude spots of opposite polarities in the same hemisphere
are generated, in agreement with what is observed e.g. in II~Peg
(e.g. Kochukhov et al., 2012). Despite of the azimuthal dynamo wave
generated by the east- or westward migration of the non-axisymmetric
modes, the solutions reported have been non-oscillatory, therefore
providing no explanation for the flip-flop type phase jumps.

In search for oscillatory solutions, one approach has been to include
small amounts of differential rotation into the models (see
e.g. Elstner \& Korhonen, 2005), essentially meaning that the dynamo
models are of $\alpha^2\Omega$ type, where differential rotation is
present, but is subdominant in the magnetig field generation. In this
way, mixed solutions with an oscillatory axisymmetric mode and a
stable non-axisymmetric one provide flip-flop type oscillatory
behavior. On the other hand, it has been known since a long time that
pure $\alpha^2$ dynamos do have oscillatory solutions (see
e.g. Baryshnikova \& Shukurov, 1987; Br\"auer \& R\"adler, 1987). In
fact, only in the case if the $\alpha$ effect is purely homogeneous and
the system unbounded, the solutions are purely
non-oscillatory. Recently, oscillatory $\alpha^2$ solutions have been
found in the spherical forced turbulence simulations of Mitra et
al. (2010) and in local Cartesian simulations of turbulent convection
(K\"apyl\"a et al., 2012a). In the next subsection, we
illustrate the possibility of oscillatory non-axisymmetric $\alpha^2$
solutions in a simple mean-field dynamo model.

\begin{figure}[t!]\begin{center}
\includegraphics[width=\textwidth]{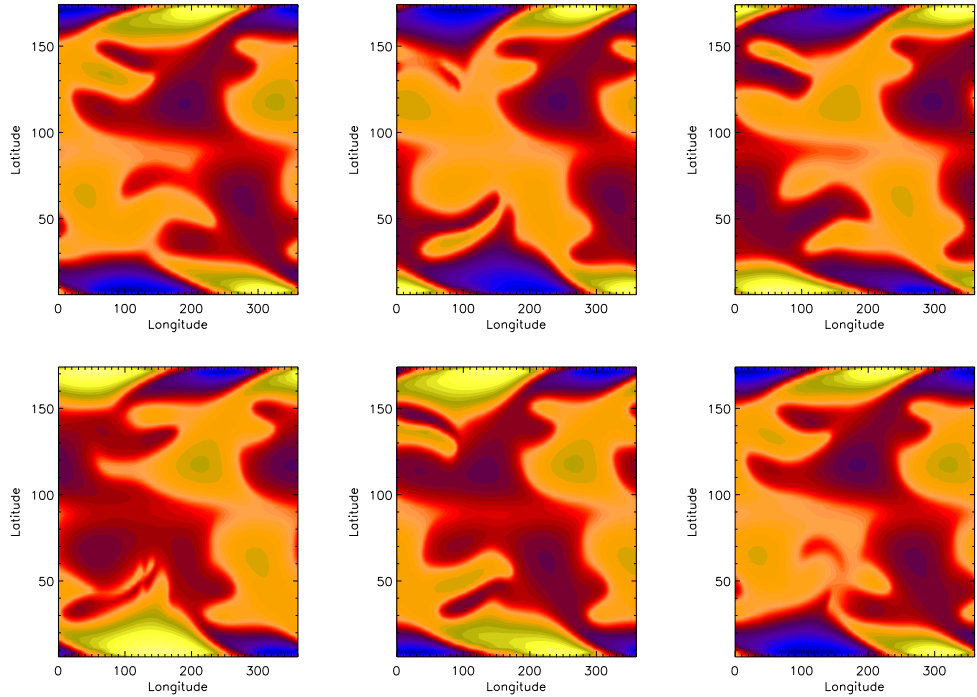}
\end{center}\caption[]{Radial magnetic field at the surface for six
  different times, $t/\tau$=(4.17,4.28,4.39,4.50,4.61,4.72) of the
  oscillation cycle of a solution with $C_{\alpha}$=$5 \times
  10^3$.}\label{fig:mfa10_cycle}\end{figure}
 
\subsubsection{Simple $\alpha^2$ dynamo model}\label{subsect:pc_dynamo}

We solve the mean-field induction equation in three dimensions using
spherical polar coordinates $(r,\theta,\phi)$. The domain is a wedge
spanning the latitude range $(6^{\circ},174^{\circ})$, radius
$0.6,1.0$ and the full azimuthal extent. The equation is solved with
the {\sc Pencil Code}\footnote{http://pencil-code.googlecode.com/}
using a mesh $32 \times 64\times 128$; the spatial discretization in
spherical polar coordinates does not allow us to explore the full
latitudinal extent, as the time step decreases strongly close to the
pole. The $\alpha$ effect has a uniform radial profile, and
cos$\theta$ dependence, obtaining maxima at the poles. The boundary
conditions applied for the magnetic field, in terms of the vector
potential that is solved for, are those of a perfect conductor at the
lower radial and latitudinal boundaries, and normal-field conditions
are applied at the top radial boundary:
\begin{eqnarray}
\frac{\pd A_r}{\pd r}= A_\theta=A_\phi =0 \,\quad \quad \quad \quad
\quad \quad \quad \quad  (r=r_1),\\
A_r=0, \frac{\pd A_{\theta}}{\pd r}=-\frac{A_{\theta}}{r}, \frac{\pd
A_{\phi}}{\pd r}=-\frac{A_{\phi}}{r}\quad\,  (r=r_3),\\
A_r=\frac{\pd A_\theta}{\pd\theta}=A_\phi=0 \quad \quad \quad \quad
\quad \quad (\theta=\theta_1,\theta_2).
\end{eqnarray}
Azimuthal boundaries are treated as periodic. As a nonlinearity, a
simple algebraic $\alpha$-quenching is used. 

\begin{figure}[t!]\begin{center}
\includegraphics[width=0.49\textwidth]{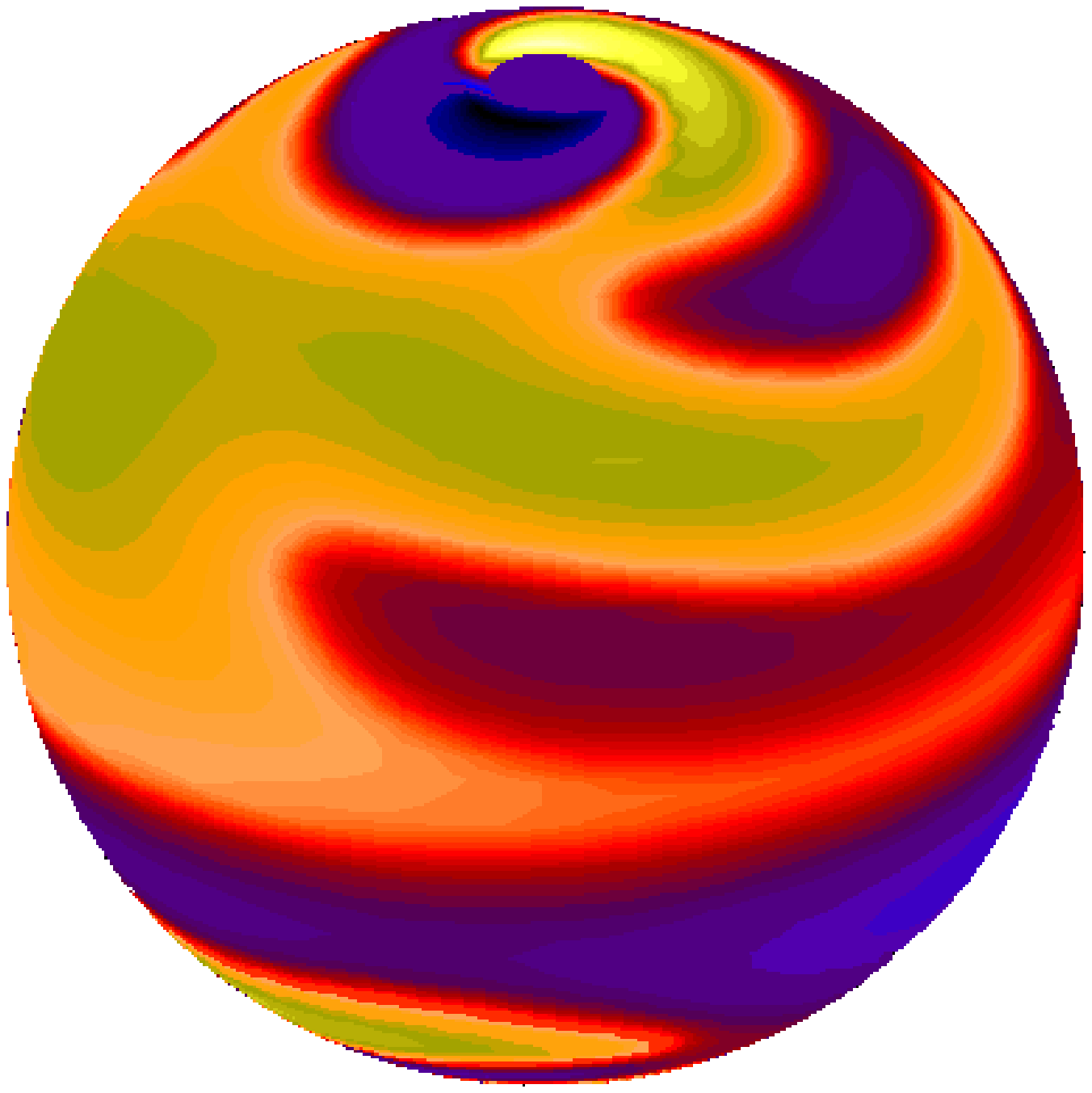}
\includegraphics[width=0.49\textwidth]{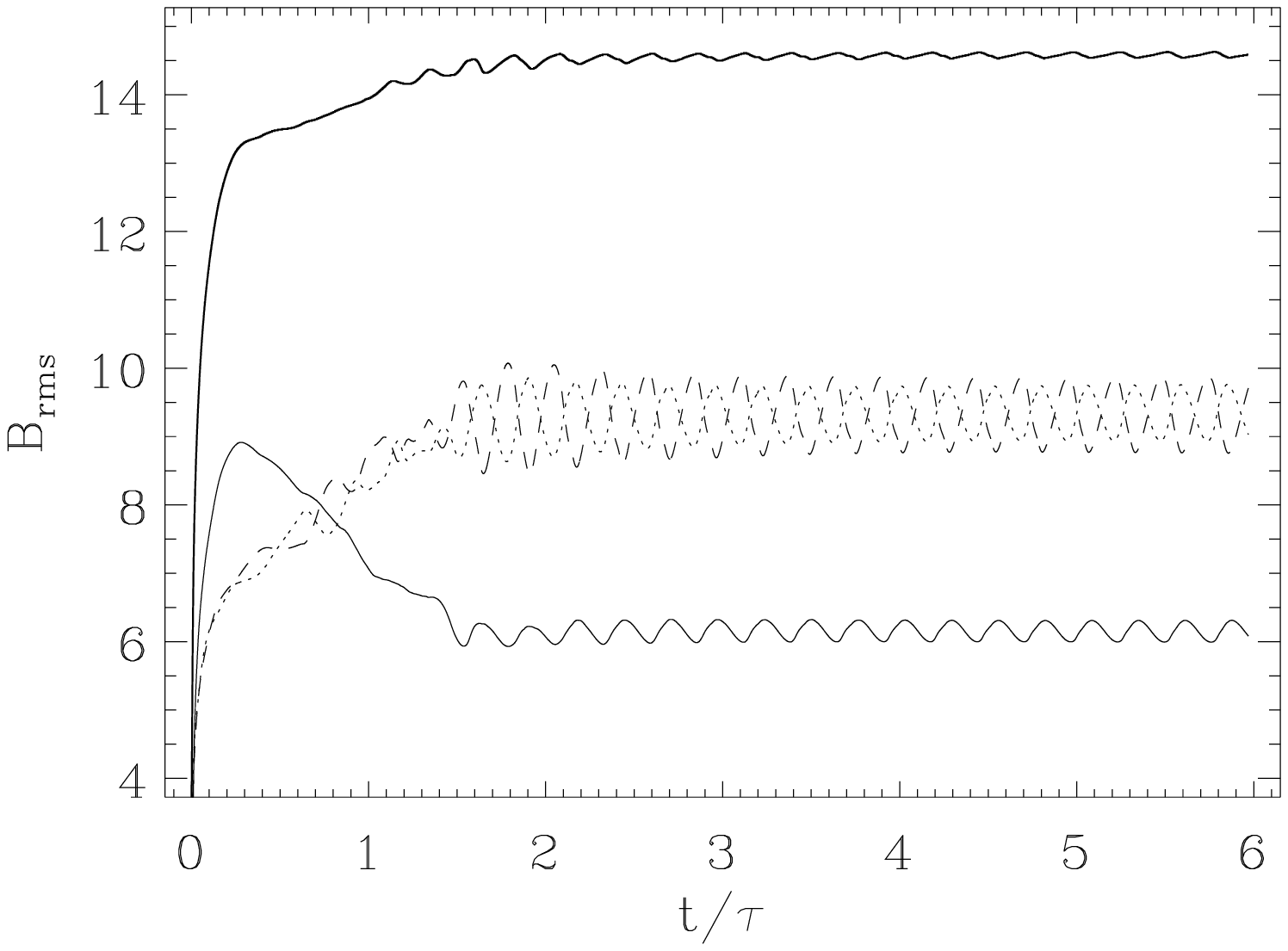}
\end{center}\caption[]{Left: snapshot of the radial magnetic field at
  the surface for $t/\tau$=4.39 in spherical projection. Right:
  magnetic energy density of the different components (total --thick
  solid, radial -- thin solid, latitudinal -- dotted, azimuthal --
  dashed).}\label{fig:mfa10}\end{figure}

We perform calculations with varying strength of the $\alpha$-effect,
spatially constant turbulent diffusion, and time being
measured in diffusion times, $\tau=d^2/\eta$, where $d$ is the depth
of the convection zone. The strength of the $\alpha$ effect is
measured by the quantity $C_{\alpha}=\alpha_0 R/\eta$, where
$\alpha_0$ is varied. The critical $C_{\alpha}$ above which growing
solutions are obtained is roughly 50. A regime of axisymmetric
oscillatory solutions of antisymmetric type is seen up to
$C_{\alpha}\approx 10^3$. With larger values, the excited solutions
become dominantly non-axisymmetric, and show eastward migration. The
solutions are otherwise non-oscillatory. When $\alpha$ effect is
ramped up further, the solutions change into
oscillatory ones. Time evolution of the radial field at the surface
during one cycle of such a calculation is shown in
Fig.~\ref{fig:mfa10_cycle}, and magnetic energy densities for
different magnetic field components in Fig.~\ref{fig:mfa10}. The
oscillation seen in the magnetic field energy densities is clearly
related to the interchange of the magnetic field strength of spots
with different polarities. Interestingly, the migration period of
eastward drift equals the cycle length.

\section{Conclusions}\label{sect:discussion}

In this paper we have discussed observations and theory of
longitudinal concentrations of solar and stellar activity
tracers. While the solar magnetic field is dominantly axisymmetric,
non-axisymmetric modes dominate the magnetic fields of rapid
rotators. The greatest challenges in the observational frontier
include understanding the uncorrelation of temperature anomalies from
magnetic field strength in ZDI, and determining whether and how the
flip-flops and phase jumps relate to stellar activity
cycles. Theoretically the most challenging task consists of explaining
how the global dynamo-generated field transforms itself into sun- and
starspots, and whether the mechanism is the same in all stars
independent of the rotation rate.

\acknowledgments The computations were performed on
the facilities hosted by the CSC -- IT Center for Science in Espoo,
Finland, which are financed by the Finnish ministry of education. We
acknowledge financial support from the Academy of Finland grant Nos.\
136189, 140970 (PJK), 218159 and 141017 (MJM), and the University of
Helsinki research project `Active Suns'. MJM and PJK thank NORDITA for
hospitality during their visits.


%
%
%
%

\begin{thebibliography}{3}

\bibitem[Augustson et al.(2012)]{ABBMT12}
Augustson, K. C., Brown, B. P., Brun, A. S., Miesch, M. S., \& 
Toomre, J.\yapj{2012}{756}{169}

\bibitem[Bai(1987)]{Bai87}
Bai, T., 1987, ApJ, 314, 795

\bibitem[Baryshnikova \& Shukurov(1987)]{BS87}
Baryshnikova, Iu \& Shukurov, A., 1987, AN, 308, 89

\bibitem[Berdyugina\&Tuominen(1998)]{BT98}
Berdyugina, S. V., \& Tuominen, I. \yana{1998}{336}{L25}

\bibitem[Berdyugina \& Usoskin(2003)]{Sveta2003}
Berdyugina, S. V., \& Usoskin, I. G.\yana{2003}{405}{1121}

\bibitem[Berdyugina \& J\"arvinen(2005)]{Sveta05}
Berdyugina, S. V. \& J\"arvinen, S. P., 2005, AN, 326, 283

\bibitem[Berdyugina et al.(1998)]{Sveta1998}
Berdyugina, S. V., Berdyugin, A. V., Ilyin, I., \& Tuominen, I.\yana{1998}{340}{437}

\bibitem[Berdyugina et al.(1998)]{Sveta1998}
Berdyugina, S. V., Berdyugin, A. V., Ilyin, I., \& Tuominen, I.\yana{1999}{350}{626}

\bibitem[Berdyugina et al.(2002)]{Sveta2002}
Berdyugina, S. V., Pelt, J. \& Tuominen, I.\yana{2002}{394}{505}

\bibitem[Berdyugina et al.(2006)]{Sveta2006} 
Berdyugina, S. V., Moss, D., Sokoloff, D., \& Usoskin, I. G.\yana{2006}{445}{703}

\bibitem[{Brandenburg \& Subramanian(2005)}]{BS05}
Brandenburg, A. \& Subramanian, K. 2005, Phys.\ Rep., 417, 1

\bibitem[Brandenburg et al.(2010)]{BKR10}
Brandenburg, A., Kleeorin, N. \&
Rogachevskii, I.\yan{2010}{331}{5}

\bibitem[Brandenburg et al.(2011)]{BKKMR11}
Brandenburg, A., Kemel, K., Kleeorin, N., Mitra, Dhrubaditya \&
Rogachevskii, I.\yapjl{2011}{740}{L50}

\bibitem[Brandenburg et al.(2012)]{BKKR12}
Brandenburg, A., Kemel, K., Kleeorin, N. \&
Rogachevskii, I.\yapj{2012}{749}{179}

\bibitem[Brown et al.(2008)]{BBBMT08}
Brown, B. P., Browning, M. K., Brun, A. S., Miesch, M. S., \& 
Toomre, J.\yapj{2008}{689}{1354}

\bibitem[Brown et al.(2011)]{BMBBT11}
Brown, B. P., Miesch, M. S., Browning, M. K., Brun, A. S., \& 
Toomre, J.\yapj{2011}{731}{69}

\bibitem[Br\"auer \& R\"adler(1987)]{BR87}
Br\"auer, H.-J., R\"adler, K.-H., 1987, AN, 308, 101

\bibitem[Choudhuri \& Gilman(1987)]{CG87}
Choudhuri, A. R. \& Gilman, P. A., 1987, ApJ, 316, 788

\bibitem[Dikpati \& Charbonneau(1999)]{DC99}
Dikpati, M. \& Charbonneau, P., 1999, ApJ, 518, 508

\bibitem[Elstner \& Korhonen(2005)]{EK05}
Elstner, D. \& Korhonen, H., 2005, AN, 326, 278

\bibitem[Ghizaru et al.(2010)]{Ghizaru10}
Ghizaru, M., Charbonneau, P., Smolarkiewicz, P. K., 2010, ApJL, 715, 133

\bibitem[Hackman et al.(2011)]{Thomas11}
Hackman, T., Mantere, M. J., Jetsu, L., Ilyin, I., Kajatkari, P., Kochukhov, O., Lehtinen, J., Lindborg, M., Piskunov, N., Tuominen, I., 2011, AN, 332, 859 

\bibitem[Hackman et al.(2012a)]{Thomas12IIPeg}
Hackman, T., Mantere, M. J., Lindborg, M., Ilyin, I., Kochukhov, O., Piskunov, N., Tuominen, I.\yana{2012a}{538}{A126}

\bibitem[Hackman et al.(2012b)]{Thomas12FKCom} Hackman, T., Pelt, J.,
  Mantere, M. J., Jetsu, L., Korhonen, H., Granzer, T., Kajatkari, P.,
  Lehtinen, J., Strassmeier, K. G., 2012b, A\&A, submitted, arXiv.org:
  1211.0914

\bibitem[Jetsu et al.(1993)]{Jetsu93}
Jetsu, L., Pelt, J., Tuominen, I.\yana{1993}{278}{449}

\bibitem[Jetsu et al.(1999)]{Jetsu99}
Jetsu, L., Pelt, J., Tuominen, I.\yana{1999}{351}{212}

\bibitem[Kitchatinov \& R\"udiger(1999)]{Kitsu99} Kitchatinov, L. L. \& R\"udiger, G.\yana{1999}{344}{911}

\bibitem[Kitiashvili et al.(2010)]{Irina10} Kitiashvili, I. N.,
  Kosovichev, A. G., Wray, A. A., Mansour, N. N., 2010, ApJ, 719, 307

\bibitem[Kleeorin et al.(1990)]{KRR90}
Kleeorin, N.I., Rogachevskii, I.V., Ruzmaikin,
A.A.\yjetp{1990}{70}{878}

\bibitem[Kochukhov et al.(2012)]{Oleg12}
Kochukhov, O., Mantere., M. J., Hackman, T., Ilyin, I., 2012, A\&A, submitted

\bibitem[Korhonen et al.(2004)]{Heidi2004}
Korhonen, H., Berdyugina, S. V., Tuominen, I., 2004, AN, 325, 402

\bibitem[Korhonen et al.(2007)]{Heidi2007}
Korhonen, H., Berdyugina, S. V., Hackman, T.,Ilyin, I. V., Strassmeier, K. G. \& Tuominen, I.\yana{2007}{476}{881}

\bibitem[K\"uker \& R\"udiger(1999)]{Kuker99} K\"uker, M. \&
  R\"udiger, G.\yana{1999}{346}{922}

\bibitem[K\"apyl\"a et al.(2011a)]{KMH11}
K\"apyl\"a, P. J., Mantere, M. J., \& Hackman, T.\yapj{2011a}{742}{34}

\bibitem[K\"apyl\"a et al.(2011b)]{KMB11}
K\"apyl\"a, P. J., Mantere, M. J., \& Brandenburg, A.\yan{2011b}{332}{833}

\bibitem[K\"apyl\"a et al.(2012a)]{KMB12a}
K\"apyl\"a, P. J., Mantere, M. J., \& Brandenburg, A., 2012a, GAFD 
(in press), arXiv:1111.6894

\bibitem[K\"apyl\"a et al.(2012b)]{KBKMR12}
K\"apyl\"a, P. J., Brandenburg, A., Kleeorin, N., Mantere, M. J.,
\& Rogachevskii, I.\ymn{2012b}{422}{2465}

\bibitem[K\"apyl\"a et al.(2012c)]{KMB12c}
K\"apyl\"a, P. J., Mantere, M. J., Brandenburg, A., 2012c, ApJL, 755,22

\bibitem[Krause \& R\"adler(1980)]{KR80}
Krause, F., \& R\"adler, K.-H.\ybook{1980}
{Mean-field magneto\-hydro\-dy\-na\-mics and dynamo theory}
{Pergamon Press, Oxford}

\bibitem[Linborg et al.(2011)]{Marjaana11}
Lindborg, M., Korpi, M. J., Hackman, T., Tuominen, I., Ilyin, I., Piskunov, N., \yana{2011}{526}{A44}

\bibitem[Linborg et al.(2013)]{Marjaana13}
Lindborg, M., Olspert, N., Pelt, J., Mantere, M. J., \& Strassmeier, K. G., 2013, A\&A, to be submitted

\bibitem[Mantere et al.(2011)]{MKH11}
Mantere, M. J., K\"apyl\"a, P. J., \& Hackman, T.\yan{2011}{332}{876}

\bibitem[Mitra et al.(2010)]{Dhruba10}
Mitra, D., Tavakol, R., K\"apyl\"a, P. J., Brandenburg, A., 2010, ApJL, 719, 1

\bibitem[Moss et al.(1995)]{Taavi95} Moss, D., Dale, D. M., Brandenburg, A. \&
  Tuominen I.,\yana{1995}{294}{155}

\bibitem[Moss(1999)]{Taavi99} Moss, D., 1999, MNRAS, 306, 300

\bibitem[Olah et al.(2006)]{Olah2006}
Ol\'ah, K., Korhonen, H., Kov\'ari, Zs., Forg\'acs-Dajka, E., \& Strassmeier, K. G.\yana{2006}{452}{303}

\bibitem[Pelt et al.(2005)]{Jaan05}
Pelt, J., Tuominen, I., Brooke, J.\yana{2005}{429}{1093}

\bibitem[Pelt et al.(2006)]{Jaan06}
Pelt, J., Brooke, J. M., Korpi, M. J., Tuominen, I.\yana{2006}{460}{875}

\bibitem[Pelt et al.(2010)]{Jaan10}
Pelt, J., Korpi, M. J., Tuominen, I.\yana{2010}{513}{48}

\bibitem[Pelt et al.(2011)]{Jaan11}
Pelt, J., Olspert, N., Mantere, M. J., Tuominen, I.\yana{2011}{535}{23}

\bibitem[Rempel et al.(2009)]{Rempel09} 
  Rempel, M., Sch\"ussler, M., Cameron, R. H., Kn\"olker, M., 2009,
  Science, 325, 171

\bibitem[R\"adler(1975)]{R75}R\"adler, K. H., 1975, MSRSL, 8, 109

\bibitem[Rogachevskii \& Kleeorin(2007)]{RK07}
Rogachevskii, I. \& Kleeorin, N.\ypre{2007}{76}{056307}

\bibitem[Stein \& Nordlund(2012)]{SN12}
Stein, R. F. \& Nordlund, \AA, 2012, ApJL, 753, 13

\bibitem[Tao et al.(1998)]{Tao98}
Tao, L., Weiss, N. O, Brownjohn, D. P., Proctor, M. R. E., 1998, ApJ, 496, 39

\bibitem[Tuominen et al.(2002)]{me02} Tuominen, I., Berdyugina, S. V. \& Korpi,
  M. J. 2002, AN, 323, 361

\bibitem[Usoskin et al.(2005)]{Usoskin05}
Usoskin, I. G., Berdyugina, S. V., \& Poutanen, J.\yana{2005}{441}{347}

\end{thebibliography}
\end{document}